\documentclass[a4paper,11pt]{article}
\pdfoutput=1
\usepackage{jhep-mod}
\usepackage[T1]{fontenc}
\usepackage{braket}
\usepackage{color,soul}
\newcommand*\dif{\mathop{}\!\mathrm{d}}

\title{Dark photon searches with atomic transitions}
\author[a]{C. \'Alvarez-Luna}
\author[a]{and J. A. R. Cembranos}
\affiliation[a]{Departamento de F\'isica Te\'orica and IPARCOS, Universidad Complutense de Madrid,\\ E-28040 Madrid, Spain}
\emailAdd{c.a.luna@ucm.es}
\emailAdd{cembra@fis.ucm.es}

\abstract{Dark matter could be made up of dark photons, massive but very light particles whose interactions with matter resemble those of usual photons but suppressed by a small mixing parameter. We analyze the main approaches to dark photon interactions and how they can be applied to direct detection experiments which test different ranges of masses and mixings. A new experiment based on counting dark photons from induced atomic transitions in a target material is proposed. This approach appears to be particularly appropriate for dark photon detection in the meV mass range, extending the constraints in the mixing parameter by up to eight orders of magnitude with respect to previous experiments.}

\arxivnumber{1812.08501}

\begin{document}
\maketitle
\flushbottom

\section{Introduction}
\subsection{Dark matter}
Many different observations strongly suggest the existence of a new kind of matter known as dark matter (DM).
Its electromagnetic interaction with baryonic matter is null, or extremely weak, and it has never been directly detected. Its existence has been deduced from analyzing gravitational effects, such as galactic rotation curves, gravitational lensing, the cosmic microwave background (CMB) and the large scale structure of the universe.
Nowadays DM is believed to account for most of the matter in the Universe, being approximately five times more abundant than baryonic matter, and it is a main ingredient of cosmological and galactic models.

A great variety of theories have tried to describe what kind of particles DM is made of, but no direct detection of such particles has been observed so far. Numerous attempts have been made to look for signals of DM beyond its gravitational
interactions in different experiments and astrophysical observations, but they have not provided any strong evidence yet.
At present, there are some popular theories such as axions or Weakly Interacting Massive Particles (WIMPs) that have been extensively studied as DM candidates. In the present work, we will focus on a different candidate known in the literature as dark photon (DP), hidden photon or paraphoton, a vector gauge boson which belongs to the class of
very weakly interacting particles (sometimes known as WISPs or Weakly Interacting Slim Particles)~\cite{DP2}.

\subsection{Dark photons}
From the point of view of a gauge theory, the DP can be described by an extension of the Standard Model containing an extra ``hidden'' $U'(1)$ symmetry: $SU(3)_C\times SU(2)_L \times U_Y(1) \times U'(1)$. All Standard Model particles are assumed to have zero charge under this additional $U'(1)$. At a theoretical level, the appearance of this type of particles is also supported by quite different models. As an example, compactifications of different string models do introduce in general new hidden $U(1)$ gauge groups (see for example~\cite{hiddenU1,hiddenU2}).

Once the electroweak symmetry is broken, this new local symmetry provides extra terms in the usual Lagrangian, resulting in something of the following form~\cite{DP1} for describing the two Abelian gauge bosons:

\begin{equation}
\mathcal{L} = -\frac{1}{4}\left(F^{\mu\nu}F_{\mu\nu}+\phi^{\mu\nu}\phi_{\mu\nu}+2\chi\phi^{\mu\nu}F_{\mu\nu}\right)-\frac{M^2}{2}\phi_\mu\phi^\mu-J_\mu A^\mu\,,
\end{equation}
where $J^\mu$ is the ordinary charged current, $A^\mu$ is the four-potential of the ordinary photon and $F_{\mu\nu}=\partial_\mu A_\nu-\partial_\nu A_\mu$ its field strength. $\phi^\mu$ and $\phi_{\mu\nu}$ are the equivalent quantities for the DP, which is endowed with a mass term $M$. In addition, a kinetic mixing term parameterized by the
constant $\chi$ is included.

In order to have DPs as viable DM candidates, we need to produce them in order to populate the universe with the proper abundance. The misalignment mechanism, which has been widely used with other candidates like axions (i.e.~\cite{mis-axions}), would result in the DPs forming a boson condensate compatible with the expected DM abundance~\cite{DP1}. This mechanism describes how the DP field, whose Compton wavelength is larger than the horizon in the early universe and whose initial value does not coincide with the minimum of the potential, acquires after inflation some spatial density which can be seen as a coherent state. In addition, the mass for the DP can arise via Higgs mechanism or Stueckelberg mechanism~\cite{Stu,Stu2}. This mass term can be very small since the (approximated) gauge symmetry stabilizes its value against radiative corrections. There are some other mechanisms proposed to produce DPs as DM with a broad range of masses (see i.e.~\cite{Raymond}). 
Particles generated this way are non-relativistic and behave as cold DM even for these tiny masses
\cite{Cembranos:2012kk,Cembranos:2012ng,Cembranos:2013dia}. In fact, although the coherent state of the DP may have a particular polarization, it has associated an isotropic average energy-momentum tensor
\cite{Cembranos:2012kk,Cembranos:2013cba} and supports isotropic cosmologies. In any case, bounds on the DP mass can be obtained from supporting a proper structure formation in the
early Universe \cite{Cembranos:2015oya, Cembranos:2016ugq}.

Besides its mass, the other important parameter for the DP phenomenology is its mixing with the ordinary photon. In fact, it is this property that will allow for its detection. For example, an explicit derivation of the mixing between two different $U(1)$ gauge groups and its physical implications can be found in~\cite{DP4}. 
Basically, this property results in two important consequences. On the one hand, the photon-DP mixing gives rise to an oscillation between both states. Some possible experiments which make use of this kind of interaction, known as ``light shining through a wall'' (LSTAW) are described in~\cite{DP2}. The most relevant ones will be summarized in Section~\ref{sec:experiments}.

On the other hand, and since the ordinary photon couples to ordinary matter, this mixing gives place to an effective interaction of the DP with matter, for example with the atoms of a detector. In Section~\ref{sec:our}, we will apply this effect to develop a specific DP detection experiment.
It can be described mathematically by rewriting the Lagrangian in terms of ``propagation eigenstates''~\cite{DP1}. This is done by applying a transformation of the following form: $A\rightarrow A-\chi \phi $ and $ \phi \rightarrow \phi+\mathcal{O}(\chi^2)$. Terms of higher order in $\chi$ can be ignored because the mixing is expected to be very small. The resulting Lagrangian is
\begin{equation}
 \mathcal{L} = -\frac{1}{4}\left(F^{\mu\nu}F_{\mu\nu}+\phi^{\mu\nu}\phi_{\mu\nu}\right)-\frac{M^2}{2}\phi_\mu\phi^\mu-J_\mu (A^\mu-\chi\phi^\mu).
\end{equation}

Written like this, we can explicitly see that the Lagrangian is diagonal in the kinetic terms for the photon and the DP, and that a coupling of $\phi^\mu$ to charged currents is present, suppressed by the mixing parameter $\chi$. The eigenstates obtained from the Lagrangian written in this form are called {\it massless} and {\it heavy} photon. It can be seen that, if we work with these eigenstates, the coupling of the heavy photon (that can be approximately identified with the DP) is proportional to the mixing parameter $\chi$ and independent of $M$. So even if $M = 0$, the heavy photon is coupled to any electromagnetic current. The only suppression of such a coupling comes from $\chi$.

From the above Lagrangian, we can also obtain another form which represents the previous behaviour (interactions between photon and DP). The eigenstates of this representation are the so called ``flavor eigenstates'': {\it interacting} and {\it sterile} photon.  We just need a rotation $\tilde{A} = A-\chi \phi $ and $\tilde{\phi} = \phi + \chi A $, with the resulting Lagrangian being
\begin{equation}
 \mathcal{L} = -\frac{1}{4}\left(\tilde{F}^{\mu\nu}\tilde{F}_{\mu\nu}+\tilde{\phi}^{\mu\nu}\tilde{\phi}_{\mu\nu}\right)-\frac{M^2}{2}(\tilde{\phi}_\mu-\chi \tilde{A}_\mu)(\tilde{\phi}^\mu-\chi \tilde{A}^\mu)-J_\mu \tilde{A}^\mu\,,
\end{equation}
where we can identify a $\tilde{\phi}-\tilde{A}$ oscillation from the non-diagonal mass term.
It must be appreciated that in all of these cases, the resultant interaction or mixing for the new photons is very weak, as is expected for any kind of DM, so direct detection experiments must be extremely sensitive.

\section{Dark photon detection experiments}\label{sec:experiments}
There have been different ways to detect DPs, which have constrained the parameters $\chi$ and $M$. Astrophysical and cosmological observations, including different properties of the early Universe, Compton evaporation and particular decay processes, impose complementary bounds on the mass and mixing parameters of the DP.
Other experiments have looked for the dark electric and magnetic fields associated with the DP (always suppressed by a factor $\chi$) or for a drift of the fine structure constant $\alpha$.
Many of these experiments are discussed, for a very wide range of masses, in~\cite{DP1}. The ones that are most relevant for the present study will be described below in more detail.

\begin{figure}[h]\centering
\includegraphics[width=0.60\textwidth]{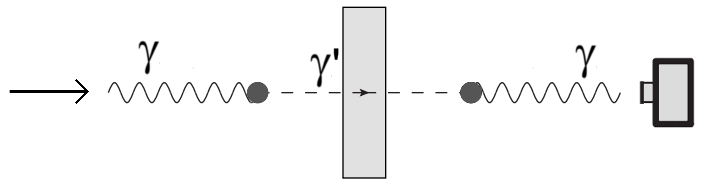}
\caption{Basic scheme of a LSTAW experiment. An incoming photon $\gamma$ is converted into a DP $\gamma'$ which interacts very weakly with the wall. It passes through the wall and is subsequently reconverted into an ordinary photon which can be detected.}
\label{fig:lstaw}
\end{figure}

Most of the laboratory searches for DPs designed up to now rely on their possible oscillation into normal photons. The first class of experiments are the ones known as ``light shining through a wall'' experiments, which basically consist in an experimental setup with different arranges of lasers, cavities and other optical devices to measure the process schematically described in figure~\ref{fig:lstaw}: 1) A beam of photons is sent into the setup, wherein one of the photons oscillates into a  DP. 2) Then this particle, which interacts very weakly with matter, can go through a layer of absorbent material; here the rest of the beam is stopped. 3) The remaining  DP oscillates back into a normal one (this process is called {\it photon regeneration}) and can be detected.

Some of these experiments are already working actively. The most significant ones are experiments like GammeV (Gamma to milli-eV particle search)~\cite{GammeV}, BMV (Bir\'efringence Magn\'etique du Vide experiment)~\cite{BMV}, LIPSS (Light Pseudoscalar and Scalar Search)~\cite{LIPSS} and ALPS (Any Light Particle Search)~\cite{ALPS}. All of them are based on the discussed procedure, varying technical details of the experimental setup. They have found different bounds for the DP mass and mixing parameter which are discussed for example in~\cite{DP2}.
This kind of laboratory experiments probes a range of parameters for masses of the order of ${\rm\ meV}$ and for mixings up to the order of $10^{-6}$. To widen this range, different detectors are needed. For instance, another option are cavity experiments, which are based on a similar principle as the previous ones, but use microwave cavities in the setup to enhance the photon-DP conversion rate. This technique results on lower bounds for the mixing ($\sim10^{-8}-10^{-12}$ is expected) and the mass ($\sim{\rm\mu eV-meV}$ or even lower)~\cite{cavities}. To this class of experiments belongs ADMX (Axion Dark Matter eXperiment)~\cite{ADMX}. Originally it was proposed to search for the conversion of axions into photons, but DPs can be detected in the same manner. Its expected sensitivity for this search reaches
$\chi \gtrsim 10^{-9}$.

Other laboratory limits on the DP mass and its mixing 
arise from other experiments like solar helioscopes which test DP production inside the Sun (such as the CERN Axion Solar Telescope, CAST~\cite{CAST}). Solar DPs propagate to the Earth, and then can oscillate into photons that can be detected by using photomultiplier tubes within a telescope. These experiments test a higher range of masses (for example, up to $1\ {\rm eV}$ with CAST). Similar reasoning but for other stars provides different types of constraints. They relate the energy ``lost'' due to DPs, which depends on the DP parameters, to the known properties of stellar evolution~\cite{HBstars}. These tests for horizontal branch stars (HB) enlarge the mass range up to $1-10^3\ {\rm keV}$, and a higher range is expected for red giant stars (RG)~\cite{constraints}. Other bounds on this mass range can be obtained from different techniques, such as multi-cathode counters~\cite{Kopylov}.

Finally, two more consequences of DP properties can be used to test a lower mass region. Firstly we have the Cavendish-type tests of the Coulomb $1/r^2$ law~\cite{Coulomb}. Massive DPs result in a modification of the potential between two charges, which can be represented by a Yukawa shape: $V(r)=\alpha\left(1+\chi^2 e^{-Mr}\right)/r$~\cite{DP2}. The correction is suppressed by $\chi^2$ so it is very small but testable, and it is more appreciable for small masses. It provides important constraints in the ${\rm \mu eV}$ and $sub-{\rm \mu eV}$ range. Several laboratory experiments test deviations of different magnitudes to obtain bounds for the DP parameters, i.e. atomic transitions and the anomalous magnetic moment of the electron~\cite{Frugiuele}.

Another test consists on analyzing the cosmic microwave background (CMB). This radiation, which was produced at the epoch of recombination with a perfect blackbody spectrum, can be distorted by the presence of DPs as they interact with background photons. This phenomenon is explained in terms of an oscillation given by the mismatch between the interaction (interacting and sterile photon) and propagation (massless and heavy photon) eigenstates. The conversion probability is given by $P_{\gamma_i\rightarrow\gamma_s} = \sin^2(2\chi) \sin^2(M^2 L/4\omega)$, where $\omega$ is the photon energy~\cite{CMB}. The oscillation length grows for smaller DP mass $M$, so the CMB provides a good probe of the low mass region being the beam-line treated, the whole universe, the longest at our disposal. Analyses of this kind have been carried out using the high precision data provided by the Far Infrared Absolute Spectrophotometer (FIRAS) on board of the COBE satellite. They lead to a bound on the mixing parameter of $\chi~\sim 10^{-7}-10^{-5}$ for a mass range between $\sim10^{-14}-10^{-7}\ {\rm eV}$. All the constraints obtained by other experiments will be later compared to the expected sensitivity of our proposal in terms of the ranges of parameter space that they cover (see figure~\ref{fig:comparison-constraints}).

\section{Dark photon detection with atomic transitions}\label{sec:our}
\subsection{Atomic transitions induced by dark photons}
In this section we describe an experimental procedure to search for DPs by using their possible coupling to matter. 
We have already discussed that other low-energy experiments have tried to detect or impose bounds on DP parameters, but here we want to describe a qualitatively different approach. The basic idea for this setup is to make use of the direct interaction of DPs with ordinary matter by trying to detect DP induced transitions in the atomic state of a chosen target atom, as has been already done in recent works, for 
example~\cite{Qiaoli}. In our case, we extend to the search of DPs the experimental setup that has been already proposed in~\cite{Sikivie} for detecting axion DM. This setup allows to adjust the different atomic transition energies involved in the detection process in order to have an optimal performance. This proposal for axions has been pursued by different experimental groups, so the aim of our work is to extend the cited study to the phenomenology of DPs. 
As we have previously discussed, the observational constraints on the DP parameters are quite wide, so an experiment for detecting them should be designed to cover a large range of the parameter space. This will be done by appropriately choosing and manipulating the atoms which compose the target.

The general form for a Lagrangian for the coupling of the DP ($\gamma'$) field to a fermion $\psi$ with charge $q$, where as usual $\bar{\psi}=\psi^\dagger\gamma^0$, and $\gamma^\mu$ are the Dirac matrices, is

\begin{equation}
\mathcal{L}_{\gamma'\bar{\psi}\psi} = -q\chi \bar{\psi} \gamma^\mu A'_\mu \psi\,,
\end{equation}
where $A'^\mu = (A'^0, \vec{A'})$ is the four-potential for the DP, being $A'^0=\phi'$. It is analogous to the interaction term for photons associated with standard QED, where $q \chi$ represents the coupling, or modified charge, of the DP to fermions. It is just $q$ for usual photons, but here it has a different value suppressed by $\chi$.

We will focus on the coupling to two possible fermions, the electron ($e$) and the nucleons ($p,n$) of the target atom.
It is possible to work in the non-relativistic limit, since DPs are assumed to be cold DM despite the value of their mass~\cite{DP1}. In any case, we can chose a semi-classical treatment for the matter-radiation interaction, represented by the DP magnetic field $B'^i = \varepsilon^{ijk}(\partial_j A'_k)$.
In general, DPs could aslo couple to matter by means of its associated electric field. However, we will focus on analyzing the magnetic field interaction because in this way we can make use of the devices designed for other DM candidates. Then, we will also be able to compare our expected sensitivity with already projected experimental setups~\cite{Sikivie}.

A complete hamiltonian for the interaction can be written as
\begin{equation}
H = -\frac{\nabla^2}{2m} + i\frac{q\chi}{2m}\left[\vec{\nabla}\cdot\vec{A}'(\vec{r},t) + \vec{A}'(\vec{r},t)\cdot\vec{\nabla}\right] + \frac{(q\chi)^2}{2m}\vec{A}'^2(\vec{r},t)-\chi \vec{\mu}\cdot\vec{B}'(\vec{r},t) + q\chi\phi'(\vec{r},t) + V(\vec{r})\,.
\end{equation}
Here, we have used a model of minimal coupling of $\chi$-suppressed electromagnetic-like radiation to a spin 1/2 particle of mass $m$, magnetic moment $\vec{\mu}$ and charge $q$~\cite{GP}, in which the relevant couplings can be represented by factors of the form

\begin{equation}
-\chi\vec{\mu} \cdot \vec{B}'= -\chi \mu_f \vec{J}\cdot (\vec{\nabla}\times\vec{A}')\,,
\end{equation}
where $\vec{J}$ is the corresponding spin and $\mu_f$ the value of the associated magnetic moment.
For a given fermion $f$, such a constant can be written in terms of the gyromagnetic factor $g_f$ for the coupling and some standard magnitude. For example, for the electron, $\mu_e = g_e \mu_B$, where $\mu_B = e/2m_e$ is the Bohr magneton and $g_e\simeq2$. On the other hand, for the nucleons $\mu_p= g_p \mu_{\text{nuc}}$ and $\mu_n = g_n \mu_{\text{nuc}}$, where $\mu_{\text{nuc}}=e/2m_p$ is the nuclear magneton~\cite{NucInt}.

It must be noted that here and from this point forward all the interactions treated are with DPs, so all the quantities $A'_i$, $B'_i$ and so on are the ones associated with them. We will apply this reasoning to the case of the interaction with an atom, which represents our detector. It will be characterized by an electron spin $\vec{S}$ and a nuclear spin $\vec{I}$, so the relevant interaction term can be written as

\begin{equation}
H = -\chi(\mu_e \vec{S} + \mu_N \vec{I}) \cdot \vec{B}'.
\end{equation}
Later, we will particularize our study to the cases of coupling to electrons or nucleons separately.

Now we want to analyze a transition in the target atom, from its ground state $\ket{0}$ (with energy $E_0$) to an excited state $\ket{i}$ (with energy $E_i$), induced by the absorption of a DP. As usual, this transition will present a resonance when the energy of the absorbed DP equals the difference between those levels: $M = E_i - E_0$.

In order to prevent the thermal population of the excited state so the inverse process $\ket{i}\rightarrow\ket{0}$ will not interfere with the measurements, the target atoms must be cooled to a temperature low enough for having all of them initially in the ground state. We can estimate the needed temperature to have almost none of the target atoms per mole in excited states. For a Maxwell distribution of particles with energy $M$ (the energy of the excited state) at a temperature $T$, this can be assured by imposing $N_A e^{-M/T} < 0.1$, and thus having less than an atom per mole initially in excited states. This requirement implies $T<0.0175\ M$, i.e.
\begin{equation}
T < 203\ {\rm mK} \left(\frac{M}{{\rm meV}}\right).
\label{eq:temperature}
\end{equation}

We can calculate the transition rate $\mathcal{R}_i :=\mathcal{R}_{\ket{0}\rightarrow \ket{i}}$ on resonance for the described process as

\begin{equation}
\label{eq:tr-rate}
\mathcal{R}_i=\frac{2\chi^2}{M} \min(t,t_i,t')\cdot\int \dif^3p\frac{\dif^3n}{\dif p^3}(\vec{p})\,\sum\limits_{\alpha=1}^3|\bra{i}(\mu_e \vec{S} + \mu_N \vec{I})\cdot (\vec{p}\times\vec{\xi}_\alpha)\ket{0}|^2.
\end{equation}
Here $\vec{\xi}_\alpha$ and $\vec{p}$ are the DP helicity and its momentum, respectively. A sum over all possible DP helicities $\alpha$ has been included to obtain an unpolarized transition rate. The function
$\dif^3n/\dif p^3$ represents the local momentum distribution~\cite{Sikivie}.

In the transition rate there are three different times involved: $t$ is the measurement integration time, set by the experimental setup; $t_i$ is the mean lifetime of the excited atomic state, fixed by the choice for the target; and $t'$ is the coherence time of the DP signal, which is defined by the associated frequency spread $t'^{-1} = \delta E/2\pi$. The energy dispersion of the DP signal is given by its mass and its average velocity squared $\bar{v^2}$
\begin{equation}
\delta E = M\left(1+\frac{1}{2} \bar{v^2}\right).
\end{equation}

In the setup the time $t$ can be adjusted, and $t'$ is inverse to the DP mass so it will usually be very large (given that $M\sim {\rm meV}$). So the constraint on the minimum time will be in general set by the atomic time $t_i$.

We find the following expression for the transition rate in Eq.~(\ref{eq:tr-rate}):

\begin{equation}
\label{eq:tr-rate2}
\mathcal{R}_i=\frac{4\chi^2}{3M} \min(t,t_i,t')\cdot\int \dif^3p \, \frac{\dif^3n}{\dif p^3}(\vec{p})\,|\vec{p}|^2\,|\bra{i}(\mu_e \vec{S} + \mu_N \vec{I})\ket{0}|^2.
\end{equation}

It is convenient to define a new dimensionless parameter $g_i$ which gives the magnitude of the effective coupling strength of the DP to the target

\begin{equation}
\label{eq:g_i}
g_i^2 \bar{v^2} M \rho\, \mu_B^2 \equiv
\int \dif^3p\,\frac{\dif^3 n}{\dif p^3}(\vec{p})\,|\vec{p}|^2 \,
|\bra{i}(\mu_e \vec{S} + \mu_N \vec{I})\ket{0}|^2\,,
\end{equation}
where $\mu_B = 9.274\cdot10^{-24}\ {\rm J/T} =  1.949\cdot10^{-22}\ {\rm s}$ is the Bohr magneton, and we have used the local DP energy density

\begin{equation}
\label{eq:rho}
\rho = M \int \dif^3p\, \frac{\dif^3 n}{\dif p^3}(\vec{p})\,.
\end{equation}

\subsection{Experimental setup}\label{sec:setup}

\begin{figure}[h]\centering
\includegraphics[width=0.65\textwidth]{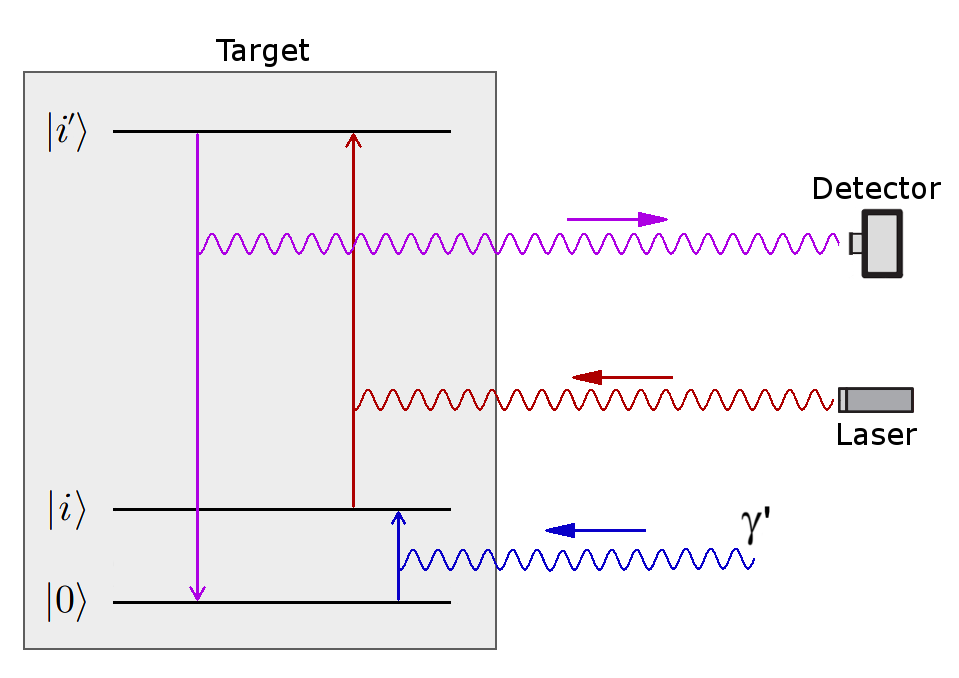}
\caption{Scheme of the experimental setup for detecting DPs.}
\label{fig:setup}
\end{figure}

Now we can describe the experimental procedure, which is based on the atomic transitions described before. The basic idea for the setup can be seen in figure~\ref{fig:setup}. The target is made of a chosen type of atoms, previously cooled to a temperature low enough (see Eq.~(\ref{eq:temperature})) for them to be in their ground state $\ket{0}$. When an incoming DP $\gamma'$ hits the target, it will induce a transition $\ket{0}\rightarrow \ket{i}$ to an excited level with the correct energy $E_i-E_0\simeq M$ (resonance condition). As usual for non-relativistic particles, the momentum contribution to the DP energy is negligible, since its velocity is very small.

After this, a new transition to a more excited state $\ket{i}\rightarrow \ket{i'}$ is induced by a properly tuned laser. This level $\ket{i'}$ will typically have an energy $\Delta E \sim 2-3\ {\rm eV}$ from the excited state $\ket{i}$ in order to be suitable for laser transitions with $\lambda \sim 600-400\ {\rm nm}$ (red$-$violet laser). We can observe that this energy gap is much larger than the resonant energy. The laser must be also tuned so it will not cause any other transition (for example, from the ground state to some other $\ket{i''}$) to prevent the measurements from having undesired noise.
Finally, the atom will decay back to the ground state $\ket{i'}\rightarrow \ket{0}$ emitting a photon that can be detected.

Since the mass of the DP, and then the resonant energy difference $E_i-E_0$, is unknown, the experiment must be designed to cover a wide range of energies. This is achieved by adjusting the energy of $\ket{i'}$ applying an external magnetic field $B$ to modify the energy gaps between atomic levels (Zeeman effect).
Then $B$ is made to vary while photons are being counted in the detector for different energies. Thus, different possibilities for the DP mass can be tested. The efficiency of this technique for counting transitions is expected between $50\%$ and $100\%$~\cite{Sikivie}. We will now derive some conditions on the number of real ``DP events'' counted if the resonance condition is reached.

We can try to estimate numerical values for the transition rate per target atom in Eq.~(\ref{eq:tr-rate2}) taking into account the definition from Eq.~(\ref{eq:g_i}).
The resultant transition rate for a mole of target atoms, obtained by considering Avogadro's constant $N_A = 6.022\cdot 10^{23}\ {\rm mol^{-1}}$ is

\begin{equation}
N_A \mathcal{R}_i=g_i^2 N_A \frac{4}{3}\chi^2\bar{v^2} \rho \mu_B^2 \min(t,t_i,t').
\label{eq:tr-mole}
\end{equation}

We can write the same expression by using typical values for the different quantities involved in Eq.~(\ref{eq:tr-mole}): $\min(t,t_i,t') \sim 1\ {\rm s}$ constrained by the atomic decay time, $\bar{v^2} \sim 10^{-6}$, and $\rho \sim 1\ {\rm GeV\, cm^{-3}}$~\cite{Sikivie}. For the mixing, we take $\chi \sim 10^{-13}$ which is our expected range of detection. In such a case,

\begin{equation}
N_A \mathcal{R}_i\simeq\frac{1.3\cdot10^3}{{\rm s}} g_i^2 \left(\frac{\chi}{10^{-13}}\right)^2\left(\frac{\rho}{{\rm GeV/cm^3}}\right)\left(\frac{\bar{v^2}}{10^{-6}}\right) \left(\frac{\min(t,t_i,t')}{{\rm s}}\right).
\label{eq:tr-mole-n}
\end{equation}

The number of events per mole of target is then obtained by multiplying the transition rate per mole in Eq.~(\ref{eq:tr-mole-n}) by the measurement integration time $t$. The bandwidth of the detector is $B_d=1/\min(t,t_i)$, and depending on its value compared to the DP frequency spread $B'=1/t'$ we can have a different number of events occurring during a tune in which the frequency is shifted by $B_d/t$. If $B' < B_d$, that is, the typical DP time is greater than the experimental setup times, events occur only during one tune, whereas events occur during $B'/B_d$ successive tunes if $B' > B_d$ (small DP time $t'$). We include a time ratio to take this into account. The final number of events per mole per tune is then

\begin{equation}
\frac{{\rm \# events}}{{\rm mole}} = t\ N_A \mathcal{R}_i\frac{\min(t,t_i)}{\min(t,t_i,t')}.
\end{equation}

We can limit the time factor if we expect the experiment to cover a reasonable frequency range per year. Here $\nu = M/2\pi$ is the DP frequency, and we assume a $30\%$ duty cycle for the setup. Then

\begin{equation}
\frac{1}{t\min(t,t_i)} = \frac{B_d}{t}  = \frac{\nu}{0.3\ {\rm yr}} = \frac{25.75\ {\rm kHz}}{{\rm s}}\left(\frac{M}{{\rm meV}}\right).
\label{eq:freq}
\end{equation}

The expected number of events per mole is independent of the time factors if Eq.~(\ref{eq:freq}) is satisfied, and given by the following expression

\begin{equation}
\frac{{\rm \# events}}{{\rm mole}} = 0.5\ g_i^2 \left(\frac{\chi}{10^{-13}}\right)^2\left(\frac{{\rm meV}}{M}\right)\left(\frac{\rho}{{\rm GeV/cm^3}}\right)\left(\frac{\bar{v^2}}{10^{-6}}\right).
\label{eq:ev-mole}
\end{equation}

In fact, Eq.~(\ref{eq:ev-mole}) represents the average number of events. The actual expected number of events $N$ in the complete target can be described by a Poisson probability distribution~\cite{Sikivie}. The probability to have an event counted in given experimental conditions is then given by $1-e^{\epsilon N}$, where $\epsilon$ is the efficiency to count events. We analyze a confidence level (C.L.) of at least $95\%$. Therefore, $N>3/\epsilon$ is needed. The total number of events can be obtained from Eq.~(\ref{eq:ev-mole}) with
$N={\rm \# moles}\cdot({\rm \# events}/{\rm mole})$. On the other hand, the number of moles can be related to the total mass of the target $M_{tar}$ and its atomic number $A$:

\begin{equation}
{\rm \# moles} = \left(\frac{M_{tar}}{{\rm g}}\right) \frac{1}{A}.
\end{equation}

Combining all these factors, we finally obtain an expected sensitivity for the parameter $g_i$ given by

\begin{equation}
g_i \simeq 2.5 \sqrt{\left(\frac{1}{\epsilon}\right)\left(\frac{A\ {\rm g}}{M_{tar}}\right)\left(\frac{10^{-13}}{\chi}\right)^2\left(\frac{M}{{\rm meV}}\right)\left(\frac{{\rm GeV/cm^3}}{\rho}\right)\left(\frac{10^{-6}}{\bar{v^2}}\right)}\,.
\label{eq:bound-g_i}
\end{equation}

\subsection{Coupling to electrons}
To estimate the search sensitivity we need to make some assumptions about the numerical values of the involved magnitudes. As said before we take the typical values $\rho = 1\ {\rm GeV/cm^{-3}}$ and $\bar{v^2} = 10^{-6}$ for the DM, whereas $\chi$ and $M$ remain variables. For the target, we assume a suitable material with $A \leq 150$ and a mass which can be cooled to an appropriate temperature $T$ given by $M_{tar} = 10^3\ {\rm g}\ (T/{\rm mK})$~\cite{Sikivie}. This temperature is defined by Eq.~(\ref{eq:temperature}), and it is proportional to the DP mass so we can estimate the total target mass available as $(M_{tar}/{\rm g}) =2.03\cdot 10^5\ (M/{\rm meV})$. Finally, we assume a counting efficiency of $60\%$, that is $\epsilon = 0.6$. It results, from Eq.~(\ref{eq:bound-g_i}), on
a sensitivity of
\begin{equation}
g_i \simeq 0.09 \left(\frac{10^{-13}}{\chi}\right).
\label{eq:numbound-g_i}
\end{equation}
In addition, since we are now studying the coupling only to electrons, in Eq.~(\ref{eq:g_i}) we can assume $\vec{I} = 0$ for the nuclei. Then, with $|\bra{i}\vec{S}\ket{0}|^2 = 1/2$, and $\int \dif^3 p \dif^3n/\dif p^3(\vec{p})\ |\vec{p}|^2 = M \bar{v^2} \rho$ we obtain

\begin{equation}
g_i = \frac{1}{\sqrt{2}} \left(\frac{\mu_e}{\mu_B}\right) = 0.708\,,
\label{eq:g_i-e}
\end{equation}
where the magnetic moment of the electron is $\mu_e \simeq \mu_B$ ($\mu_e = 1.0011597 \mu_B$).
By combining the results from Eq.~(\ref{eq:g_i-e}) with those of Eq.~(\ref{eq:numbound-g_i}) we get

\begin{equation}
\chi \simeq 1.22\cdot 10^{-14}.
\end{equation}
This sensitivity is indicated by the shaded area in figure~\ref{fig:comparison-constraints}.

Analogously to the above discussion, we could study the coupling to nuclei, for example, by assuming $\vec{S} = 0$ for electrons. However, medium effects are expected to be more important for the nuclei-DP interaction. Note that the typical Compton wavelength of these light DM candidates is larger or comparable to the size of the atom. So, screening effects from electrons can be non-negligible.  In any case, if one performs the study, the coupling to nuclei gives a much worse sensitivity ($\chi \simeq 10^{-11}$) compared with the electron analysis. Therefore we will not pursue this study. 

\subsection{Comparison with other experiments}
\begin{figure}[h]\centering
\includegraphics[width=1\textwidth]{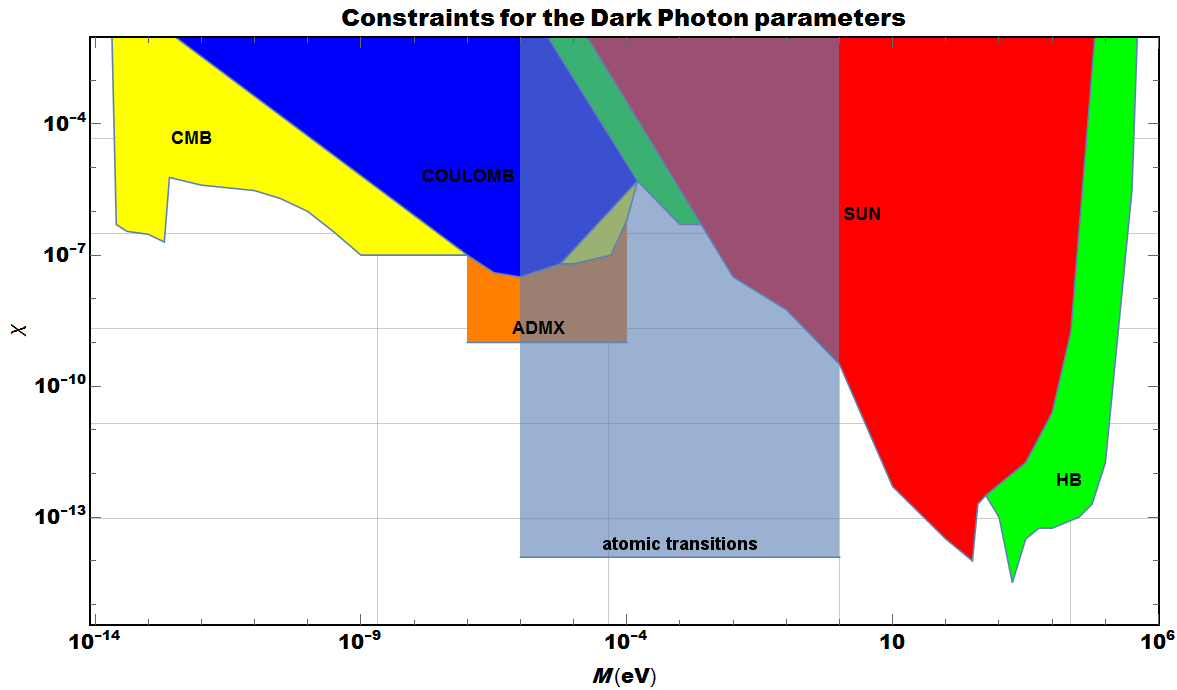}
\caption{
Expected sensitivity of the proposed detector for the coupling of the DP to electrons (shaded region), estimated at $95\%$ C.L. under the assumptions spelled out in the text. It is compared with present constraints from other analyses, in particular the other colored regions show the bounds imposed by solar (red) and horizontal branch stars (green) studies, Coulomb law (blue) and CMB (yellow) tests, and the limits that can be achieved by the ADMX experiment (orange).}
\label{fig:comparison-constraints}
\end{figure}

We can compare our results with those of other DP detection experiments presented in Section~\ref{sec:experiments}. The bounds imposed by those experiments are shown in figure~\ref{fig:comparison-constraints}. The constraints coming from LSTAW experiments 
would cover a limited region around $M\sim10^{-3}\ {\rm eV}$ and $\chi\sim10^{-6}$, which is also mostly covered by stellar constraints (solar and horizontal branch).

From figure~\ref{fig:comparison-constraints}, we observe that in the medium-mass range the proposed experiment can cover a much wider range for the mixing parameter. For  $M\sim {\rm meV}$, the most important constraints are those obtained by LSTAW experiments. From~\cite{DP2} 
we can see that the lowest bound would be the one expected for ALPS, $\chi\sim10^{-6}-10^{-7}$. For lower masses ($M<10^{-4}\ {\rm eV}$) the bounds could be improved by ADMX, whose expected sensitivity is $\chi=10^{-9}$, and for higher masses ($M>10^{-1}\ {\rm eV}$) a similar sensitivity is expected for solar helioscopes. As we have estimated, with our setup, much lower limits could be achieved following our approach, $\chi\sim10^{-14}$. Therefore, the accessible mixing range is extended by $6$ orders of magnitude from ADMX and solar bounds, and by $7-8$ orders of magnitude from light shining through a wall experiments.

\section{Conclusions and future insight}
The DP is theoretically motivated to be a viable DM candidate. It can be described by its mass $M$ and mixing parameter $\chi$. DPs have a very weak interaction with matter which resembles that of usual photons but always suppressed by $\chi$. Given that this parameter has a very low value (typical values treated in the literature are about $10^{-3}-10^{-16}$) this interaction is clearly weaker than any of the standard forces. Nevertheless, it allows for possible DP detection.

In this work, the fundamental DP interactions used for developing experimental detection have been discussed. The photon-DP oscillation effect which can be tested with optical devices is the basis of LSTAW, photon regeneration and similar experiments. This work has analyzed a different possibility for detecting DPs in the meV range of masses. The idea relies essentially on setting a target whose atomic levels are suitable for atomic transitions due to the absorption of a DP, that is, a level whose energy difference from the ground state is in resonance with the DP mass. With the proposed setup this energy difference can be conveniently tuned by applying an external magnetic field to the target (Zeeman effect), so it can be modified to probe different DP masses. In addition, these changes should not have any influence on the data collection, since the events are counted by means of an optical display (laser and photon detector) which can be easily adjusted.

For DPs expected in the ${\rm meV}$ mass range, the temperature needed for the target to have an admissible detection efficiency is of $T\sim 0.1\, {\rm K}$. In fact, temperatures as low as $T\sim 10^{-4}\, {\rm K}$ are challenging but achievable, what determines the lower DP mass detectable with this technique within the ${\rm \mu eV}$ scale. This also allows for target masses up to one kilogram, which is better to enhance detection rates given that the number of events that take place in the target is proportional to the target mass. The large range of DP masses that can be proved with this analysis is determined by the field approach we are assuming. This treatment demands an occupation number much larger than one. If we take into account a typical momentum for the DP of $p_{\rm typical}\sim 10^{-3} M$, the phase-space density can be estimated as
    \begin{equation}
        \frac{\rho}{M (p_{\rm typical})^3}\sim\left( \frac{10\, {\rm eV}}{M}  \right)^4\,,
    \end{equation}
where the local mass density is $\rho\sim 1\,{\rm GeV cm}^{-3}\sim 10^{-5}\, {\rm eV}^4$. Therefore, the occupation number is much larger than one for $M< 1\, {\rm eV}$~\cite{Guth}.

By assuming typical values for the involved magnitudes, we expect roughly $0.5$ events per mole of target during one tune. In addition, we can see that with this approach there is no dependence of the sensitivity for the mixing parameter on $M$. On the one hand, the transition rate does not depend on $M$ (Eqs.~\ref{eq:tr-rate2}, \ref{eq:g_i}, \ref{eq:rho}); and on the other hand, $\min(t, t_i)$ is inverse to $M$ (Eq.~\ref{eq:freq}). This implies that the number of events per mole is inverse to $M$, but the number of moles is proportional to this mass because the target needs to be cooled to a temperature that is proportional to $M$. Thus the DP mass cancels and the sensitivity for the mixing parameter does not depend on it at first order.

Given that and assuming a detection efficiency of $60\%$, it is expected to increase the search sensitivity to much lower values of $\chi$ in relation to other experimental techniques. As has been discussed, if the DP couples to the detector electrons, we obtain an estimation for the sensitivity of $\chi\sim10^{-14}$. This result means an improvement of $7-8$ orders of magnitude with respect to previous experiments. In figure~\ref{fig:comparison-constraints}, it can be observed how the proposed setup would cover an untested mixing range in the ${\rm meV}$ range, complementing the sensitivity of LSTAW experiments (upper mixing range), ADMX (lower mass region) and SUN+HB tests (upper mass region).

The present work describes the general outline for designing an experimental setup to detect DPs. The theoretical basis and expected results are well-motivated. Nevertheless, further study of experimental particularities must be done to develop a real setup. In the first place, detailed studies of target atoms are needed to look for the most suitable atom for the experiment. The chosen medium should be as inactive as possible, minimizing its effects on the described setup.

We can offer a brief insight on possible materials that could be used for devising a functional target. A first proposal is molecular oxygen. It was proposed for a similar device to detect axions in~\cite{oxygen}. It has some possible transitions about $10^{-3}\ {\rm eV}$, and can be cooled up to $280\ {\rm mK}$. However, molecules in a gas have undesired interactions that lead to spurious detection that cannot be controlled so in principle we reject the use of fluids.
Then, we focus on crystals. This introduces a new difficulty because the internal structure of each material must be examined in detail. The first class of crystals to be considered are those based on transition metal ions. Within these samples, absorption lines of several ${\rm meV}$ appear as a result of antiferromagnetic resonance for low temperatures without applied magnetic fields~\cite{FeF2,MnF2,CoCl2}. The presence of such fields produce splittings of $0.01 - 1\ {\rm meV}$. 
The second class of crystals to be considered are those based on rare earth ions. In them, the absorption spectra are similar to those of the pure ions and thus exhibit lines and splittings of the same order. The experimental information regarding these compounds is more recent, and they have been also considered for similar setups (i.e. axion detector such as~\cite{Sikivie}). In any case, the experimental data are very limited, with very few measurements done for each material, in specific conditions and usually for very small samples~\cite{Gd2CuO4,TmFeO3,HoFeO3}. 
Nevertheless, these results are quite encouraging because all of these materials exhibit energy levels in the desired range, and most of them respond in a promising way to magnetic fields and low temperature conditions. In addition, although the discussed materials are based on transition metal and rare earth ions, the proposed target has a crystal structure that minimizes its conductivity, so we do not expect a large background associated to this effect. In any case, a deeper research on the properties of these crystals is necessary in order to find the appropriate material for the target. 

In the second place, a complete description of an optical system to detect transitions in the target must be planned. In that matter, it must be determined which are the best components to use for the laser, detector and the remaining elements of the setup. It is also necessary to evaluate if these experimental components could cause difficulties in certain detection ranges or modify the expected theoretical sensitivity of the setup. Improvements in the experimental technology can also provide better sensitivities for the experiment, for example by increasing the efficiency of the counting technique or the duty cycle.

Finally, another possibility of this technique must be remarked. The described development can be used for other kind of DM particles. As an example, an analysis for an axion search is described in~\cite{Sikivie}, and it could be extended to other models. Thus, deeper research on experimental systems and target characteristics may result useful for searching different DM candidates.

\acknowledgments
We would like to thank A. Aranda, D. Castellanos, E. Vazquez-Jauregui, P. Amore and the entire AXILA Collaboration 
for fundamental discussions about the experimental setup.
This work has been supported by the MINECO (Spain) projects FIS2014-52837-P, FIS2016-78859-P(AEI/FEDER, UE),
and Consolider-Ingenio MULTIDARK CSD2009-00064.

\bibliographystyle{JHEP}

\end{document}